# Frailty or Frailties: Exploring Frailty Index Subdimensions in the English Longitudinal Study of Ageing

**Authors:** Lara Johnson1, Bruce Guthrie1, Paul A T Kelly1, Atul Anand1,2, Alan Marshall1, Sohan Seth1,3

**Provenance:** 1. Advanced Care Research Centre, University of Edinburgh; 2. Centre for Cardiovascular Science, University of Edinburgh, 3. School of Informatics, University of Edinburgh

**Acknowledgements**: We thank Professor Paul Kelly, who generously provided invaluable patient and public involvement inputs that greatly enriched the depth and relevance of this research.

**Funding declaration**: This research was funded by the Legal & General Group (research grant to establish the independent Advanced Care Research Centre at University of Edinburgh). The funder had no role in conduct of the study, interpretation or the decision to submit for publication. The views expressed are those of the authors and not necessarily those of Legal & General.

## ABSTRACT

**Background:** Frailty, a state of increased vulnerability to adverse health outcomes, has garnered significant attention in research and clinical practice. Existing constructs aggregate clinical features or health deficits into a single score. While simple and interpretable, this approach may overlook the complexity of frailty and not capture the full range of variation between individuals.

**Methods:** Exploratory factor analysis was used to infer latent dimensions of a frailty index constructed using survey data from the English Longitudinal Study of Ageing (ELSA), wave 9. The dataset included 58 self-reported health deficits in a representative sample of community-dwelling adults aged 65+ ($N$ = 4971). Deficits encompassed chronic disease, general health status, mobility, independence with activities of daily living, psychological wellbeing, memory and cognition. Multiple linear regression examined associations with CASP-19 quality of life scores.

**Results:** Factor analysis revealed four frailty subdimensions. Based on the component deficits with the highest loading values, these factors were labelled "Mobility Impairment and Physical Morbidity", "Difficulties in Daily Activities", "Mental Health" and "Disorientation in Time". The four subdimensions were a better predictor of quality of life than frailty index scores.

**Conclusions:** Distinct subdimensions of frailty can be identified from standard index scores. A decomposed approach to understanding frailty has potential to provide a more nuanced understanding of an individual's state of health across multiple deficits.

**KEY MESSAGES**

- A single frailty score may overlook the complexity and possible multidimensionality of frailty and not capture the full range of variation between individuals.

- Four subdimensions of frailty were determined by factor analysis of a frailty index: (1) Mobility Impairment and Physical Morbidity, (2) Difficulty in Daily Activities, (3) Mental Health and (4) Disorientation in Time.

- Frailty subdimensions were a better predictor of CASP-19 quality of life scores than frailty index scores.

- Taking into account subdimensions of frailty has the potential to provide a more nuanced understanding of an individual's state of health across multiple deficits using existing data.

## INTRODUCTION

Rapidly ageing populations challenge healthcare systems internationally because of the increasing number of people living with multimorbidity and frailty [1]. Frailty describes a state of increased vulnerability to adverse health outcomes for individuals compared to their peers of the same age [2]. Frailty is associated with accelerated functional decline, higher mortality, increased hospital admissions, long-term care stays, primary and secondary care costs and lower quality of life [3, 4, 5].

There are several approaches to frailty assessment, including the phenotype model (defined by the presence of clinical features) [6], the Rockwood Clinical Frailty Scale (based on functional ability) [7] and the cumulative deficit model, which counts health deficits to calculate a frailty index as the proportion of possible deficits present in an individual [8]. The frailty index is a widely used instrument for identifying frail older adults from secondary data and can be constructed from both electronic health records and survey data [9]. A higher frailty index score is associated with increased risk of experiencing an adverse health event and may be categorised into levels of severity [10]. A distinctive feature of the cumulative deficit model of frailty is that frailty index scores can be automatically calculated using existing data, such as electronic health records [10]. This makes it possible to evaluate (or at least screen for) the presence of frailty across entire populations [3].

These existing constructs all represent frailty with a single score or binary status. The advantage of this approach is its simplicity. Frailty scores are a straightforward and easy-to-understand measure of frailty. The frailty index, in particular, is easy to compute (adding up counts) and does not require weights to be calibrated across different populations.

However, frailty is a complex syndrome that encompasses a range of physical, cognitive, and psychosocial factors [11]. There can be considerable variation between individuals with the same frailty index score. These variations are especially apparent between different demographic groups. Women are more frail than men [12] but live longer [13]. Women have a lower risk of mortality, even at the same frailty index scores [12], but have a higher risk of falls [14] and hip fractures [15]. This suggests that a single score may overlook the complexity and possible multidimensionality of frailty and not capture the full range of variation between individuals.

The identification of frailty subdimensions was recognised as a research priority in 2006 [16]. Since then, however, few studies have explored subdimensions of frailty. One [17], two [18] and three [19] dimensions of the phenotype model of frailty have variously been suggested. Seven factors of frailty indicators [20] and ordinal multi-morbidity items [21] have been proposed.

This study aimed to explore empirically observed patterns in health deficits to identify distinct subdimensions of frailty and assess them for their ability to explain quality of life. Quality of life was selected as the outcome measure due to its alignment with the concerns and priorities of older individuals, as underscored by our public contributor, and its established relevance in prior research on frailty in community-dwelling older people [5].

## METHODS

### Data Source

The study population was drawn from the 7289 participants in Wave 9 of the English Longitudinal Study of Ageing (ELSA), a population-based study of community-dwelling adults aged 50 years and over living in England [22]. Participants provided informed consent before participation and ethical approval was obtained from the South Central – Berkshire Research Ethics Committee (ref 17/SC/0588). Wave 9 is the most recent data available in ELSA (collected in 2018 and 2019). We excluded participants aged <65 years (*N* = 2314) and with missing data for more than 20 deficits (*N* = 4), resulting in a final sample of 4971.

### Measures

We used the frailty index developed in a previous study [23] for use in ELSA, which is comprised of 58 deficits across a range of chronic diseases (e.g., hypertension), general health (e.g., vision impairment), mobility (e.g., getting up from a chair), activities of daily living (e.g., ability to get dressed), psychological wellbeing (e.g., sadness) and memory (Supplementary Materials, Table 1). We inspected each of these deficits against the standard criteria for creating a frailty index [9] to explore population prevalence, saturation by age group and correlation with increasing age.

We used CASP-19 to assess quality of life [24]. CASP-19 is a self-report measure developed for older adults and comprises 19 items spanning control, autonomy, self-realisation and pleasure (CASP-19) [25], with composite scores ranging from 0 to 57 and higher scores representing better quality of life.

## Statistical Analysis

We performed Exploratory Factor Analysis (EFA) to determine how many factors are necessary to explain the data adequately and to estimate the loadings of each variable on the latent factors. We used the factor loadings to interpret the meaning of each factor based on the variables with high loadings, to name the factors, and to consider the implications from a clinical perspective.

Given the binary nature of the deficits, we first computed the correlations between every set of two binary variables using the Phi Correlation Coefficient [26]. After confirming that the correlation matrix was factorisable, it was used as input to EFA. Parallel analysis and the visual scree test were used to determine the appropriate number of factors to retain [27]. Parsimony and theoretical meaningfulness were also considered. Acceptable model fit is indicated by a Root Mean Square of the Residuals (RMSR) of 0.05 or less. We employed an oblimin rotation, as it was assumed that the factors would be correlated. The criteria for determining factor adequacy were established a priori. Factor loadings > 0.20 or < -0.20 were considered salient. Factors with a minimum of three salient pattern coefficients and that were theoretically meaningful were considered adequate. We conducted Pearson's correlation analysis to examine the relationships between the factor scores as well as with the frailty index score.

Two multiple linear regression models were fitted to examine their effectiveness in explaining the association between frailty and quality of life. Both models included age and sex (0 = male, 1 = female) as independent variables and the CASP-19 score as the dependent variable. Additional independent variables were the frailty index score in Model 1 and the four factor scores in Model 2. All variables were standardised. We compared the two models' $R^2$ scores.

Analyses were performed using R (version 4.3.1) [28] and its *psych* package (Version 2.3.3) [29]. Missing data were handled using pairwise deletion [30].

## RESULTS

The study population comprised 4971 people aged 65 years and older (Table 1), of which 56.6% of were female. The mean age was 74.9 years (SD = 7.0). 95.4% of respondents had at least 1 deficit. The mean number of deficits was 7.68 (SD = 6.96) and the mean frailty score was 0.13 (SD = 0.12). The mean CASP-19 score was 42.64 (SD = 8.11).

The most common deficit experienced by all participants was arthritis (48.8%) and the least was having a diagnosis of Alzheimer's disease (1.0%) (Supplementary Materials, Table 2). None of the deficits were universally prevalent at any age. The deficit with the highest prevalence in a single age group was difficulty climbing several flights of stairs without resting, which was present in 80.3% of respondents aged over 90. Of the 58 deficits, 53 were moderately or strongly correlated with age ($r > 0.25$). The mean correlation with age was $r = 0.66$. The deficit most correlated with age was climbing several flights of stairs ($r = 0.92$). The deficits with weak or negative correlations with age were lung disease ($r = 0.04$), asthma ($r = -0.27$), restless sleep ($r = -0.38$), pain whilst walking ($r = -0.44$) and having any psychiatric condition ($r = -0.67$).

Parallel analysis and visual inspection of the scree plot suggested that four factors should be retained. Through factor analysis with the oblimin rotation method, we extracted four factors. The four factors had eigenvalues of 10.9 (12%), 3.4 (9 %), 2.4 (5%) and 1.9 (2%) respectively. The RMSR was 0.02, indicating acceptable model fit.

The factor loadings of the deficits, which represent the strength and direction of the relationship between observed variables and the latent factors, are shown in Figure 1.

Factor 1 was saliently loaded by 25 items, including difficulty climbing several flights of stairs without resting (0.73), kneeling (0.69), lifting 10 pounds (0.69), getting up from a chair (0.67), walking for 100 metres (0.64) and performing various other movements. It also emphasised the burden of chronic diseases such as arthritis (0.48), hypertension (0.24), and osteoporosis (0.20). We labelled this factor "Mobility Impairment and Physical Morbidities", given that the largest loadings were for a range of physical mobility limitations and because health-related issues exhibited stronger loadings on this factor compared to any other factor.

Factor 2 was saliently loaded by 18 items, including difficulty taking medications (0.77), managing money (0.75), cooking (0.72), making phone calls (0.69), having a diagnosis of dementia (0.48) and poor eyesight (0.27). Items included basic self-care, such as difficulty using the toilet (0.49) and bathing (0.47) as well as instrumental activities of daily living (IADLs), such as difficulty shopping for groceries (0.48). IADLs are higher-level activities that individuals typically need to perform to live independently and that generally involve cognitive and organisational skill in addition to physical fitness. Arthritis, which had a strong positive loading on Factor 1 (0.48), had a low negative loading (-0.23) on this factor. Restless sleep (-0.12) negatively loaded onto this factor. We named this factor "Difficulties in Daily Activities" because it primarily encompassed items related to difficulties in performing a wide range of activities.

Factor 3 was saliently loaded by eight items, including depression (0.72), sadness (0.60), feelings of loneliness (0.50) and sleep disturbances (0.26). We labelled this factor "Mental Health".

Factor 4 was saliently loaded by eight items, including incorrectly recalling the year (0.55), the month (0.54) and the day of the month (0.45). Dementia loaded more highly onto Factor 2 (0.48) than this factor (0.28). Factor 4 had several negative loadings, including difficulty using the toilet (-0.26), getting in and out of bed (-0.25) and eating (-0.21). We named this factor "Disorientation in Time".

The factors were not mutually exclusive, and an individual can have a high score on more than one frailty subdimension. The Pearson correlations between factor scores ranged from 0.118 (Factor 3 and Factor 4) to 0.664 (Factor 1 and Factor 2). All factor scores correlated positively with the frailty index score (Figure 2), ranging from $r$ = 0.294 for Factor 4 to $r$ = 0.929 for Factor 1.

The diversity of frailty subdimensions can be illustrated by comparing two women in their early 80s with an identical frailty index score of 0.24. This single score denotes a moderate level of frailty but considering the frailty subdimensions reveals distinct profiles (Figure 3): one displays an elevated "Mental Health" score, whereas the other presents a heightened "Mobility Impairment and Physical Morbidities" score.

The results of two multiple linear regressions models (Figure 3) revealed that factor scores (Model 2) were a better predictor of quality of life than frailty index scores (Model 1). Model 2 exhibited a higher explanatory power ($R^2 = 0.3538$, adjusted $R^2 = 0.3527$, p = < .001) compared to Model 1 ($R^2 = 0.3008$, adjusted $R^2 = 0.3002$, p = < .001).

In Model 2, sex (β = 0.19, t = 7.01, p < .000) and Factor 2 (β = 0.08, t = 4.12, p < .001) were positively associated with quality of life, signifying that being female and higher Factor 2 scores were linked to increased quality of life. Factor 3 (β = -0.40, t = -26.28 p < .000), Factor 1 (β = -0.37, t = -19.70, p < .001), age (β = -0.08, t = -6.16, p < .001) and Factor 4 (β = -0.04, t = -2.35, p < .01) exhibited negative associations. In substantive terms, this implies that a one standard deviation increase in Factor 1 corresponds to a 0.37 standard deviation reduction in quality of life.

## DISCUSSION

Our factor analysis of the standard frailty index in ELSA identified four subdimensions: “Mobility Impairment and Physical Morbidities”, “Difficulties in Daily Activities”, “Mental Health” and “Disorientation in Time”. The correlation between these subdimensions was low, except for a moderate association between “Mobility Impairment and Physical Morbidities” and “Difficulties in Daily Activities”. Individuals’ overall frailty index score was largely driven by these factors, with “Mental Health” and “Disorientation in Time” showing a weaker association and suggesting that these subdimensions may not be sufficiently highlighted in a single frailty score. The four subdimensions outperformed frailty index scores in explaining associations with quality of life.

The strengths of this study are the large number of participants in a representative sample of the population and the availability in ELSA of a wide range of potential deficits including rich information on function, wellbeing and mental health, which are less robustly captured in routine health data datasets. “Mental Health” in particular appears differentiated from the other subdimensions observed in this study, with significant loadings *only* for psychological symptoms, depression, overall general health and the presence of psychiatric conditions. “Difficulties in Daily Activities” captures a different set of difficulties than physical mobility impairments, which could be a useful distinction. There is precedent for using a frailty index on ELSA data to predict mortality [31], assess determinants of frailty [32] and examine cohort differences in levels and trajectories of frailty [33].

The study has several limitations. Firstly, the reliance on a single data source may limit generalisation to other populations, such as care home residents not included in ELSA. Secondly, the cross-sectional nature of the study restricts the ability to discern temporal patterns or causal relationships. Integrating and interpreting data from multiple dimensions can be challenging, as it requires considering the interrelationships and potential interactions among the different dimensions.

The existing literature offers different perspectives on the dimensionality of frailty (Supplementary Materials, Table 4). Two studies examined Fried's phenotype model, proposing one [17] and two [18] subdimensions. However, the utility of factor analysis may be limited, given that the phenotype model consists of only five components. Another study [19] evaluated the dimensionality of the 15-item Groningen Frailty Indicator (GFI) questionnaire, which incorporates various frailty indicators but does not include specific diseases. The study proposed three dimensions: Daily Activities, Psychosocial Functioning and Health Problems [19].

Two studies explored the dimensionality of a frailty index [20, 21]. One study [20] explored the dimensionality of 35 binary frailty indicators in a UK sample of 4286 women aged 60-79 years, while the other used factor analysis on 30 ordinal multimorbidity items in a Canadian sample of 649 adults [21]. Both studies identified seven distinct factors, though there were variations in the nature of these factors. Four of the factors consistently aligned across the studies: Cardiac Symptoms, Respiratory Symptoms, Psychological Problems [20] / Emotional Wellbeing [21], and Co-Morbidities. Physical Ability and Physiological Measures [20] approximately corresponded to Physical Activity and Mobility [21]. However, Visual Impairment [20] and Instrumental Health [21] were distinct.

Our study discovered four subdimensions that explained the relationship among 58 frailty index items. Only one dimension closely aligned with previously documented factors in the literature. "Mental Health" corresponded to Psychosocial functioning [19], Psychological Problems [20], and Emotional Wellbeing [21]. This convergence of findings across different studies underscores the consistency and relevance of the Mental Health dimension in frailty research.

Two of our subdimensions share some similarities with factors from previous studies but also exhibit significant differences. “Mobility Impairment and Physical Morbidities” amalgamates several dimensions identified in prior literature: Mobility [21], Physical Activity [21], Cardiac Symptoms [20] [21], Respiratory Symptoms [20] [21], Health Problems [19] and Co-Morbidities [20] [21], indicating a close interconnection between physical mobility and morbidities. “Difficulties in Daily Activities” resembles the GFI “Daily Activities” subscale [19], but includes IADLs as well as basic self-care.

One of our subdimensions is entirely different. “Disorientation in Time” was not encountered in previous studies, even those that incorporated items on memory complaints [19] or problems [20].

While quality of life generally worsens with frailty [5], there are variances. An advantage of using frailty subdimensions is that the contribution of various aspects of frailty to quality of life can be unpicked. The positive association between female sex with improved quality of life underscores that frailty outcomes manifest differently in men and women. “Mental Health” and “Mobility Impairment and Physical Morbidities” were identified as strong contributors to reduced quality of life, having a larger effect than increasing age. The small positive association between “Difficulties in Daily Activities” and quality of life challenges conventional assumptions about the role of IADLs on overall wellbeing. The small negative association of “Disorientation in Time” indicates that memory issues have a relatively minor influence on quality of life.

The presence of numerous items within a frailty index presents an ideal opportunity for consolidating into subdimensions. However, it is crucial to acknowledge that the specific items included in frailty indices can vary widely across different studies and populations (Supplementary Materials, Table 5). The absence of a single, uniform frailty index means that results may not be directly transferable from one study to another. This underscores the need for caution and rigor when comparing and generalising results across studies.

An important next step in advancing our understanding of frailty subdimensions is to replicate this work by examining frailty indices constructed from different data sources, including routine electronic health records. Future research should also focus on investigating the implications of the identified factor structure of frailty for adverse health outcomes. Understanding how frailty subdimensions relate to specific health outcomes, such as hospitalisations, falls, fractures and mortality, would provide valuable information on the differential prognostic profiles associated with each frailty subdimension.

Taking into account subdimensions of frailty has the potential to provide a more nuanced understanding of an individual's state of health across multiple deficits than a single frailty score does. A decomposed approach to understanding frailty has the potential to be more meaningful to clinicians and patients, but further research is required to replicate findings and to understand associations with outcomes.

Table 1: Study Population Characteristics

| | Men | | Women | | Total | |
|---|---|---|---|---|---|---|
| | **N** | **%** | **N** | **%** | **N** | **%** |
| **N** | 2156 | 43.4% | 2815 | 56.6% | 4971 | |
| **Age (Years)** | | | | | | |
| 65-69 | 563 | 26.1% | 731 | 26.0% | 1294 | 26.0% |
| 70-79 | 1052 | 48.8% | 1293 | 45.9% | 2345 | 47.2% |
| 80-89 | 483 | 22.4% | 670 | 23.8% | 1153 | 23.2% |
| 90+ | 58 | 2.7% | 121 | 4.3% | 179 | 3.6% |
| **Number of Deficits** | | | | | | |
| 0 | 122 | 5.7% | 108 | 3.8% | 230 | 4.6% |
| 1-5 | 1145 | 53.1% | 1178 | 41.8% | 2323 | 46.7% |
| 6-10 | 464 | 21.5% | 679 | 24.1% | 1143 | 23.0% |
| 11-15 | 200 | 9.3% | 376 | 13.4% | 576 | 11.6% |
| 16+ | 225 | 10.4% | 474 | 16.8% | 699 | 14.1% |
| **Number of Deficits, mean (SD)** | 6.56 (6.29) | | 8.54 (7.31) | | 7.68 (6.96) | |
| **Frailty index score, mean (SD)** | 0.113 (0.109) | | 0.147 (0.126 ) | | 0.132 (0.119) | |
| **CASP-19 score, mean (SD)** | 42.80 (7.90) | | 42.50 (8.27) | | 42.64 (8.11) | |

Figure 1: Factor Loadings

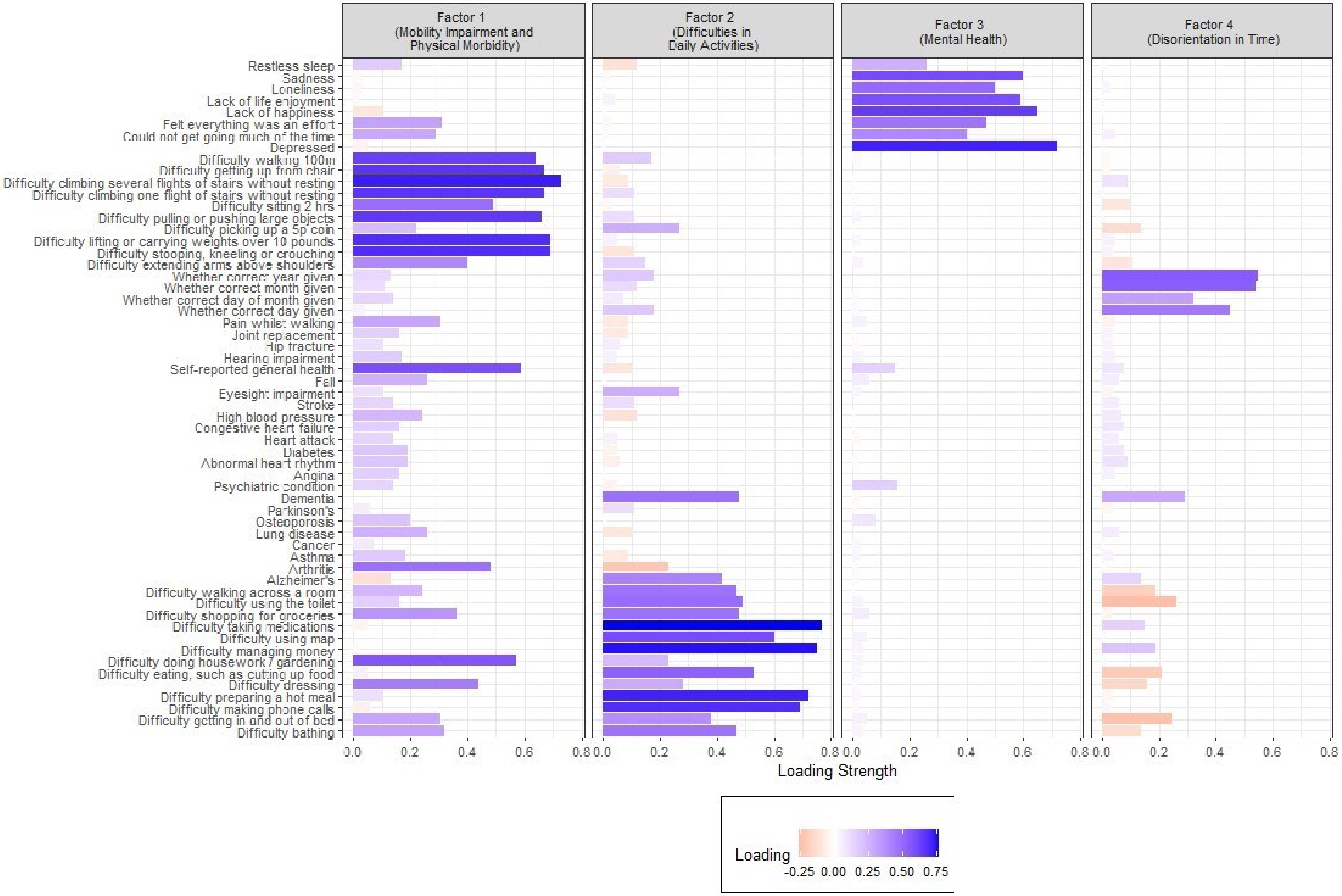

Figure 2: Comparison of Individuals' Latent Factor Scores and Frailty Index Scores, illustrating the variation in frailty subdimension for each frailty index score

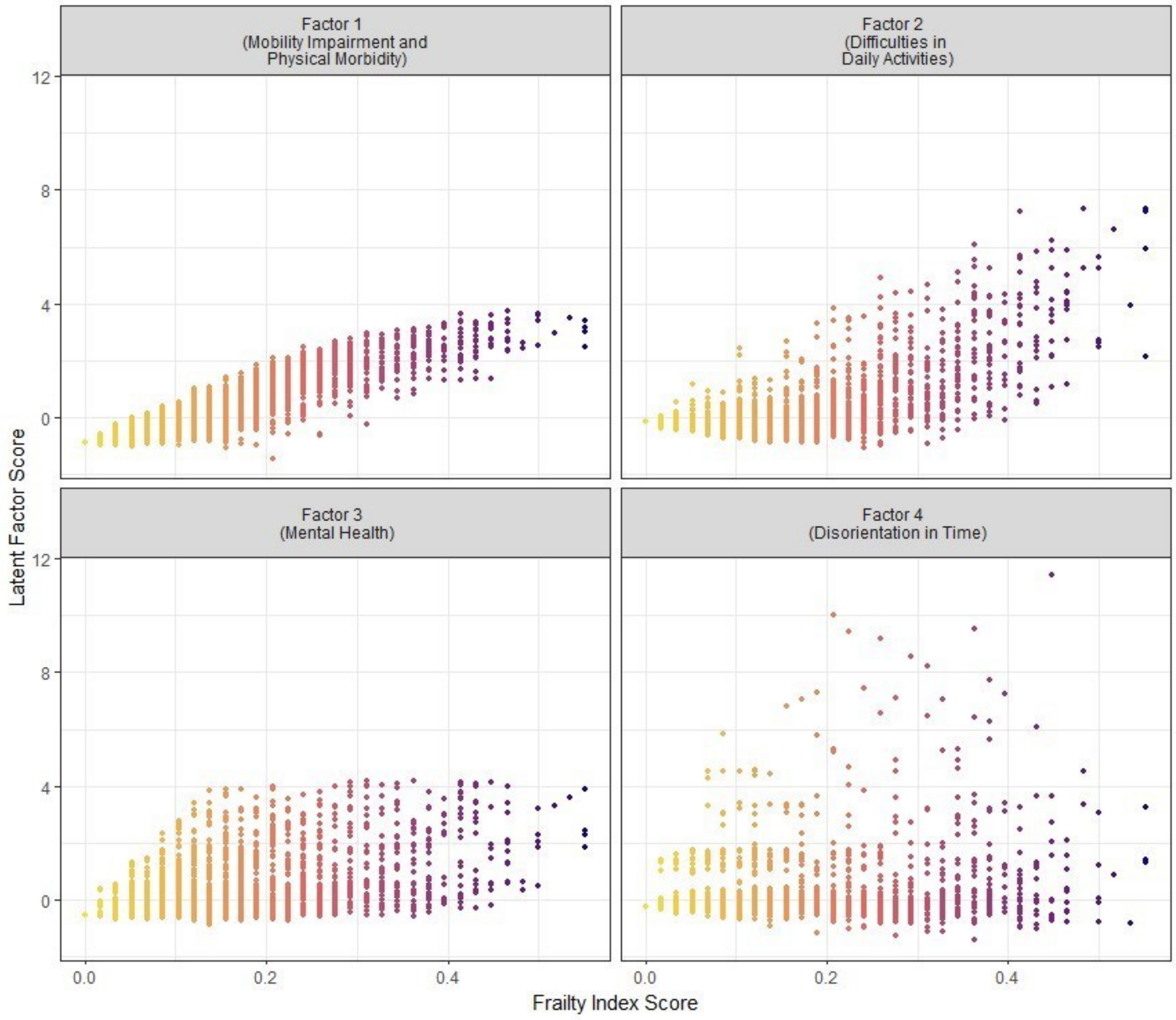

Figure 3: Illustration of frailty subdimensions in two individuals with the same frailty index score (0.24), matched for age group (80-85) and sex (F)

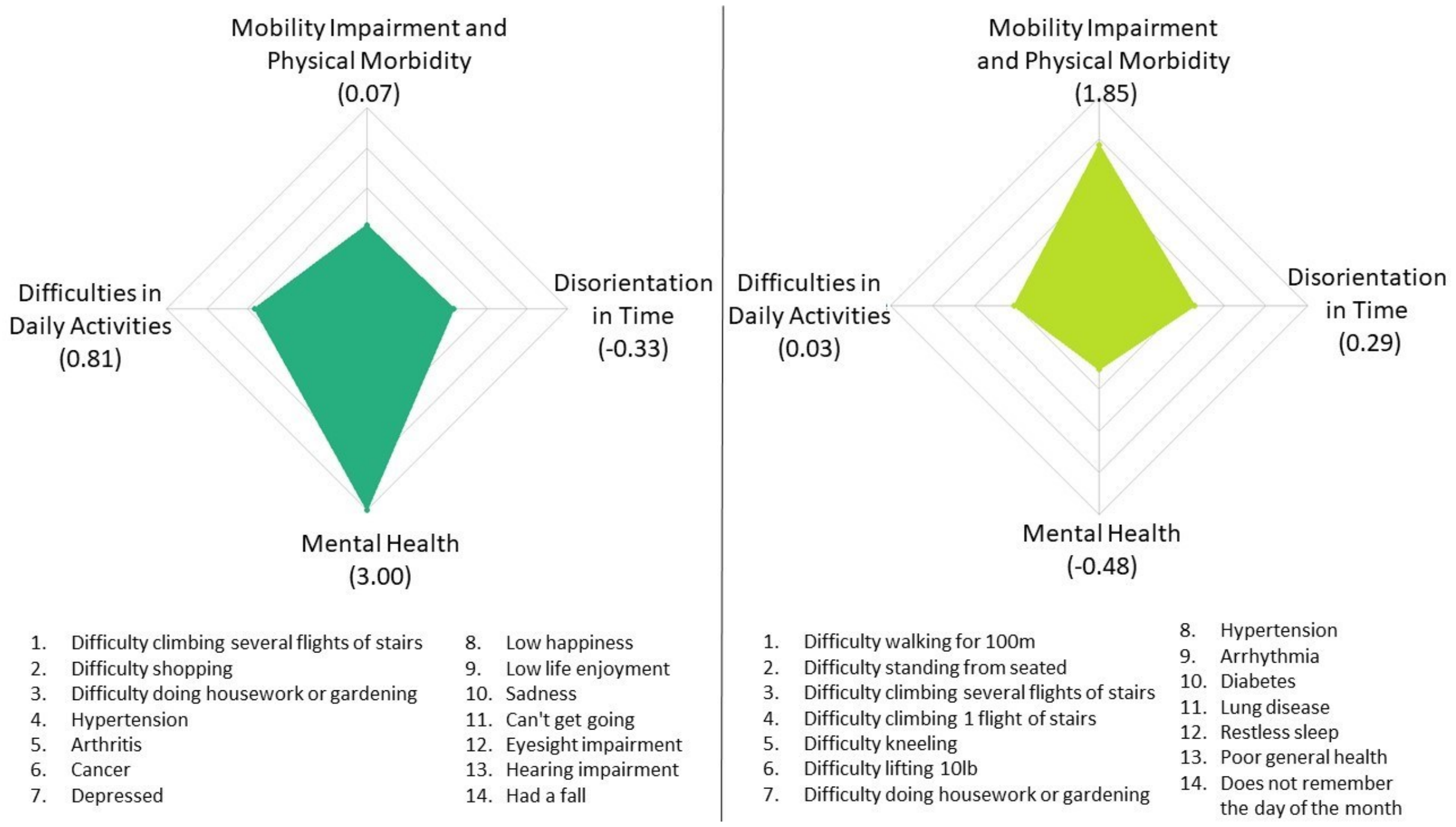

Figure 4. Forest Plot of Two Multiple Linear Regression Models Examining the Association between Frailty Index Scores (Model 1) and Factor Scores (Model 2) with Quality of Life

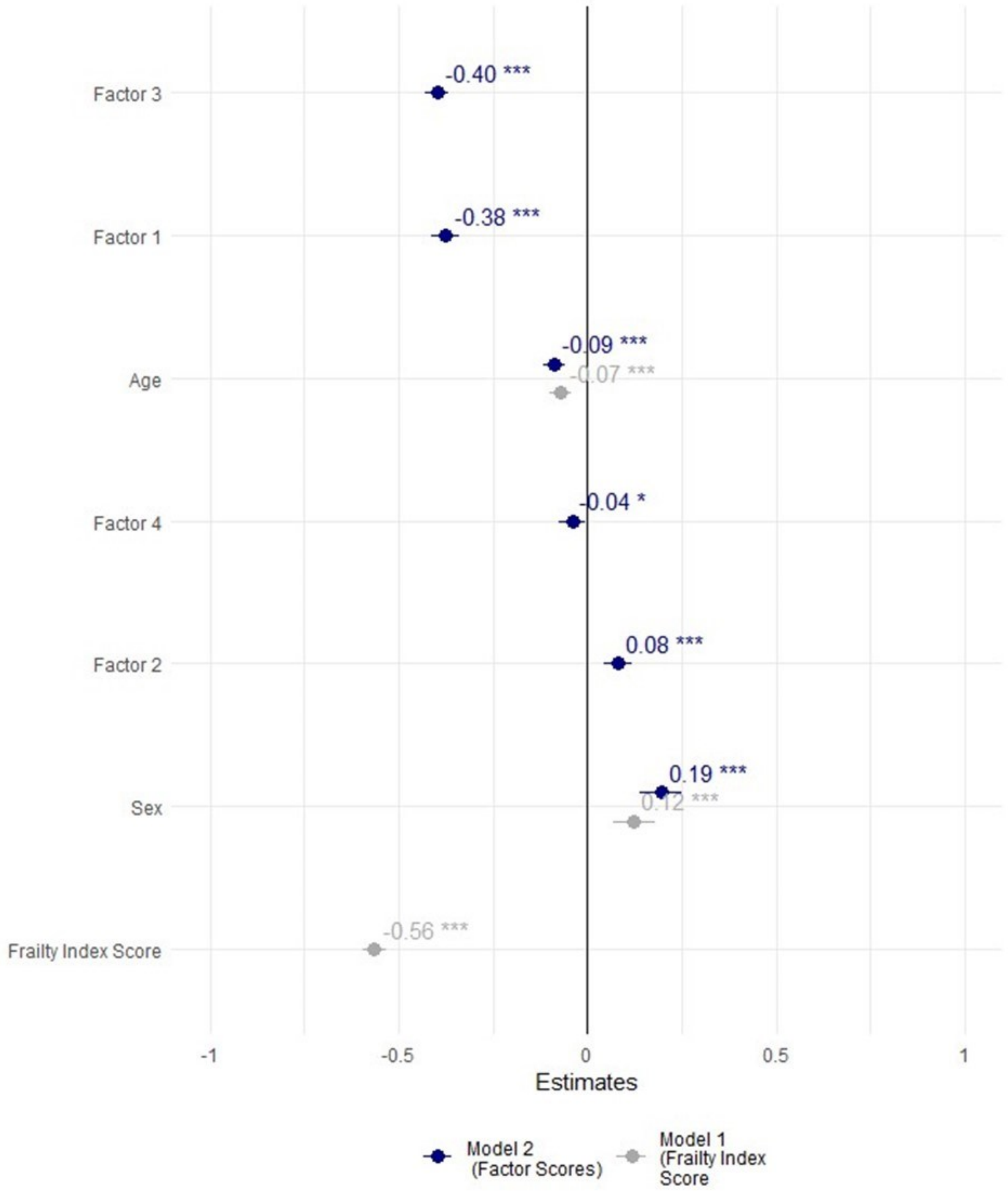

## ETHICS STATEMENTS

**Patient consent for publication**

Not application.

**Ethics approval**

The English Longitudinal Study of Ageing (ELSA) Wave 9 received ethical approval from the South Central – Berkshire Research Ethics Committee on 10th May 2018 (17/SC/0588). Participants gave informed consent to participate in the study before taking part.

## ACKNOWLEDGEMENTS

We thank Professor Paul Kelly, who generously provided invaluable patient and public involvement inputs that greatly enriched the depth and relevance of this research.

## FUNDING DECLARATION

This research was funded by the Legal & General Group (research grant to establish the independent Advanced Care Research Centre at University of Edinburgh). The funder had no role in conduct of the study, interpretation or the decision to submit for publication. The views expressed are those of the authors and not necessarily those of Legal & General.

## Supplementary Material

Table 1: List of Deficits

| ELSA Variable Name | Description |
|---|---|
| hemobwa | Difficulty walking 100m |
| hemobsi | Difficulty sitting 2 hrs |
| hemobch | Difficulty getting up from chair |
| hemobcs | Difficulty climbing several flights of stairs without resting |
| hemobcl | Difficulty climbing one flight of stairs without resting |
| hemobst | Difficulty stooping, kneeling or crouching |
| hemobre | Difficulty extending arms above shoulders |
| hemobpu | Difficulty pulling or pushing large objects |
| hemobli | Difficulty lifting or carrying weights over 10 pounds |
| hemobpi | Difficulty picking up a 5p coin |
| headldr | Difficulty dressing |
| headlwa | Difficulty walking across a room |
| headlba | Difficulty bathing |
| headlea | Difficulty eating, such as cutting up food |
| headlbe | Difficulty getting in and out of bed |
| headlwc | Difficulty using the toilet |
| headlma | Difficulty using map |
| headlpr | Difficulty preparing a hot meal |
| headlsh | Difficulty shopping for groceries |
| headlph | Difficulty making phone calls |
| headlme | Difficulty taking medications |
| headlhg | Difficulty doing housework / gardening |
| headlmo | Difficulty managing money |
| hedimbp | High blood pressure |
| hediman | Angina |
| hedimmi | Heart attack |
| hedimhf | Congestive heart failure |
| hedimar | Abnormal heart rhythm |
| hedimdi | Diabetes |
| hedimst | Stroke |
| hediblu | Lung disease |
| hedibas | Asthma |
| hedibar | Arthritis |
| hedibos | Osteoporosis |
| hedibca | Cancer |
| hedibpd | Parkinson's |
| hedibps | Psychiatric condition |
| hedibad | Alzheimer's |
| hedibde | Dementia |
| psceda | Depressed |
| pscedb | Felt everything was an effort |
| pscedc | Restless sleep |
| pscedd | Lack of happiness |
| pscede | Loneliness |
| pscedf | Lack of life enjoyment |
| pscedg | Sadness |
| pscedh | Could not get going much of the time |
| hehelf* | Self-reported general health |
| heeye* | Eyesight impairment |
| hehear* | Hearing impairment |
| hefla | Fall |
| hefrac | Hip fracture |
| heji | Joint replacement |
| mmpain | Pain whilst walking |
| cfdatd | Whether correct day of month given |
| cfdatm | Whether correct month given |
| cfdaty | Whether correct year given |
| cfday | Whether correct day given |
| cfmem | Memory prompt needed |
| cflisenq | Whether reaches cutoff for cflisen |
| cfaniq | Whether reaches cutoff for cfani |
| cflisdq | Whether reaches cutoff for cflisd |
| cflisen | Number of words recalled immediately |
| cfani | Number of animal mentioned |
| cflisd | Number of words recalled after delay |

Unless otherwise indicated, deficits were binary (0 = no deficit is reported; 1 = deficit is reported).

*Ordinal variables were dichotomised from the reported 5-point Likert scale (Excellent, Very good, Good, Fair, Poor), where Fair or Poor were treated as deficit present).

Table 2: Prevalence of Deficits

| ELSA Variable Name | Deficit Description | Prevalence | | |
|---|---|---|---|---|
| | | Men | Women | Total |
| hedibar | Arthritis | 39.0% | 56.4% | 48.8% |
| hedimbp | High blood pressure | 46.4% | 44.6% | 45.4% |
| hemobst | Difficulty stooping, kneeling or crouching | 38.1% | 50.3% | 45.0% |
| pscedc | Restless sleep | 33.0% | 45.8% | 40.3% |
| hemobcs | Difficulty climbing several flights of stairs without resting | 28.7% | 45.6% | 38.3% |
| hefla | Fall | 26.0% | 29.8% | 28.1% |
| hemobch | Difficulty getting up from chair | 23.3% | 29.9% | 27.1% |
| hehelf | Self-reported general health | 25.9% | 27.3% | 26.7% |
| hemobli | Difficulty lifting or carrying weights over 10 pounds | 15.8% | 33.9% | 26.1% |
| hehear | Hearing impairment | 31.0% | 21.3% | 25.5% |
| hemobpu | Difficulty pulling or pushing large objects | 13.0% | 25.4% | 20.0% |
| headlhg | Difficulty doing housework / gardening | 16.0% | 22.0% | 19.4% |
| pscedh | Could not get going much of the time | 15.4% | 21.7% | 18.9% |
| pscedb | Felt everything was an effort | 14.5% | 21.9% | 18.7% |
| hemobcl | Difficulty climbing one flight of stairs without resting | 13.4% | 21.4% | 17.9% |
| pscedg | Sadness | 10.9% | 21.5% | 16.9% |
| hemobwa | Difficulty walking 100m | 14.3% | 17.9% | 16.4% |
| cfdatd | Whether correct day of month given | 17.4% | 15.5% | 16.3% |
| headldr | Difficulty dressing | 15.5% | 15.3% | 15.4% |
| hedimdi | Diabetes | 16.8% | 12.3% | 14.3% |
| hedibos | Osteoporosis | 4.5% | 18.5% | 12.4% |
| hemobsi | Difficulty sitting 2 hrs | 9.7% | 13.9% | 12.1% |
| hemobre | Difficulty extending arms above shoulders | 8.7% | 14.3% | 11.9% |
| headlsh | Difficulty shopping for groceries | 8.4% | 14.1% | 11.7% |
| hedibas | Asthma | 10.1% | 12.8% | 11.6% |
| headlba | Difficulty bathing | 9.6% | 12.3% | 11.2% |
| hedimar | Abnormal heart rhythm | 12.1% | 9.7% | 10.7% |
| pscede | Loneliness | 6.8% | 13.4% | 10.5% |
| psceda | Depressed | 6.6% | 13.4% | 10.5% |
| hedibca | Cancer | 10.6% | 9.4% | 10.0% |
| hedibps | Psychiatric condition | 5.3% | 9.9% | 7.9% |
| hemobpi | Difficulty picking up a 5p coin | 7.3% | 8.0% | 7.7% |
| mmpain | Pain whilst walking | 5.7% | 9.2% | 7.7% |
| pscedf | Lack of life enjoyment | 4.4% | 9.9% | 7.5% |
| pscedd | Lack of happiness | 5.0% | 9.3% | 7.5% |
| headlbe | Difficulty getting in and out of bed | 5.8% | 8.6% | 7.4% |
| headlpr | Difficulty preparing a hot meal | 5.5% | 8.2% | 7.0% |
| hediblu | Lung disease | 7.7% | 6.0% | 6.8% |
| headlma | Difficulty using map | 4.7% | 7.9% | 6.5% |
| headlmo | Difficulty managing money | 5.3% | 5.8% | 5.6% |
| headlwa | Difficulty walking across a room | 4.5% | 5.6% | 5.1% |
| headlwc | Difficulty using the toilet | 3.5% | 6.4% | 5.1% |
| hedimst | Stroke | 5.8% | 4.3% | 4.9% |
| headlme | Difficulty taking medications | 4.7% | 4.4% | 4.5% |
| headlph | Difficulty making phone calls | 5.0% | 3.8% | 4.3% |
| heji | Joint replacement | 3.6% | 4.8% | 4.3% |
| heeye | Eyesight impairment | 3.4% | 4.7% | 4.2% |
| hediman | Angina | 4.1% | 3.2% | 3.6% |
| hedimmi | Heart attack | 4.5% | 2.7% | 3.5% |
| headlea | Difficulty eating, such as cutting up food | 2.9% | 3.8% | 3.4% |
| cfdaty | Whether correct year given | 3.3% | 3.3% | 3.3% |
| hedibde | Dementia | 2.8% | 3.1% | 3.0% |
| cfdatm | Whether correct month given | 2.9% | 2.8% | 2.9% |
| cfday | Whether correct day given | 1.5% | 2.6% | 2.1% |
| hedimhf | Congestive heart failure | 1.6% | 1.2% | 1.4% |
| hedibpd | Parkinson's | 1.8% | 0.8% | 1.2% |
| hefrac | Hip fracture | 1.1% | 1.1% | 1.1% |
| hedibad | Alzheimer's | 1.1% | 1.0% | 1.0% |

Table 3: Factor Loadings

| Deficit | Factor 1 | Factor 2 | Factor 3 | Factor 4 |
|---|---|---|---|---|
| Difficulty walking 100m | 0.64 | 0.17 | -0.03 | -0.03 |
| Difficulty sitting 2 hrs | 0.49 | -0.03 | 0.01 | -0.10 |
| Difficulty getting up from chair | 0.67 | -0.06 | 0.00 | -0.03 |
| Difficulty climbing several flights of stairs without resting | 0.73 | -0.09 | 0.01 | 0.09 |
| Difficulty climbing one flight of stairs without resting | 0.67 | 0.11 | -0.01 | 0.01 |
| Difficulty stooping, kneeling or crouching | 0.69 | -0.11 | -0.01 | 0.04 |
| Difficulty extending arms above shoulders | 0.40 | 0.15 | 0.04 | -0.11 |
| Difficulty pulling or pushing large objects | 0.66 | 0.11 | 0.03 | -0.01 |
| Difficulty lifting or carrying weights over 10 pounds | 0.69 | 0.05 | 0.03 | 0.05 |
| Difficulty picking up a 5p coin | 0.22 | 0.27 | 0.01 | -0.14 |
| Difficulty dressing | 0.44 | 0.28 | 0.02 | -0.16 |
| Difficulty walking across a room | 0.24 | 0.47 | -0.01 | -0.19 |
| Difficulty bathing | 0.32 | 0.47 | 0.04 | -0.14 |
| Difficulty eating, such as cutting up food | 0.05 | 0.53 | 0.03 | -0.21 |
| Difficulty getting in and out of bed | 0.30 | 0.38 | 0.05 | -0.25 |
| Difficulty using the toilet | 0.16 | 0.49 | 0.04 | -0.26 |
| Difficulty using map | 0.00 | 0.60 | 0.05 | 0.01 |
| Difficulty preparing a hot meal | 0.10 | 0.72 | 0.03 | -0.04 |
| Difficulty shopping for groceries | 0.36 | 0.48 | 0.06 | -0.04 |
| Difficulty making phone calls | -0.06 | 0.69 | -0.02 | 0.03 |
| Difficulty taking medications | -0.05 | 0.77 | 0.00 | 0.15 |
| Difficulty doing housework / gardening | 0.57 | 0.23 | 0.04 | -0.01 |
| Difficulty managing money | -0.02 | 0.75 | 0.04 | 0.19 |
| High blood pressure | 0.24 | -0.12 | 0.01 | 0.07 |
| Angina | 0.16 | -0.01 | 0.00 | 0.05 |
| Heart attack | 0.14 | 0.05 | -0.03 | 0.06 |
| Congestive heart failure | 0.16 | 0.00 | 0.01 | 0.08 |
| Abnormal heart rhythm | 0.19 | -0.06 | 0.02 | 0.09 |
| Diabetes | 0.19 | -0.05 | 0.00 | 0.08 |
| Stroke | 0.14 | 0.11 | 0.00 | 0.06 |
| Lung disease | 0.26 | -0.10 | 0.00 | 0.06 |
| Asthma | 0.18 | -0.09 | 0.03 | 0.04 |
| Arthritis | 0.48 | -0.23 | 0.02 | 0.02 |
| Osteoporosis | 0.20 | -0.01 | 0.08 | 0.00 |
| Cancer | 0.07 | 0.00 | 0.03 | -0.01 |
| Parkinson's | 0.06 | 0.11 | -0.02 | -0.04 |
| Psychiatric condition | 0.14 | -0.05 | 0.16 | -0.01 |
| Alzheimer's | -0.13 | 0.42 | 0.02 | 0.14 |
| Dementia | 0.00 | 0.48 | -0.03 | 0.29 |
| Depressed | -0.05 | 0.01 | 0.72 | -0.01 |
| Felt everything was an effort | 0.31 | -0.03 | 0.47 | 0.00 |
| Restless sleep | 0.17 | -0.12 | 0.26 | -0.02 |
| Lack of happiness | -0.10 | 0.01 | 0.65 | -0.01 |
| Loneliness | 0.03 | -0.01 | 0.50 | 0.03 |
| Lack of life enjoyment | -0.02 | 0.04 | 0.59 | 0.01 |
| Sadness | -0.03 | -0.03 | 0.60 | 0.00 |
| Could not get going much of the time | 0.29 | -0.02 | 0.40 | 0.05 |
| Self-reported general health | 0.59 | -0.10 | 0.15 | 0.08 |
| Eyesight impairment | 0.10 | 0.27 | 0.03 | -0.04 |
| Hearing impairment | 0.17 | 0.05 | 0.04 | 0.05 |
| Fall | 0.26 | -0.01 | 0.06 | 0.06 |
| Hip fracture | 0.10 | 0.06 | -0.02 | 0.04 |
| Joint replacement | 0.16 | -0.09 | -0.01 | 0.04 |
| Pain whilst walking | 0.30 | -0.09 | 0.05 | -0.05 |
| Whether correct day of month given | 0.14 | 0.07 | 0.02 | 0.32 |
| Whether correct month given | 0.11 | 0.12 | 0.00 | 0.54 |
| Whether correct year given | 0.13 | 0.18 | 0.00 | 0.55 |
| Whether correct day given | 0.04 | 0.18 | 0.02 | 0.45 |

Table 4: Literature Summary

| Paper | Sarkisian et al (2008) [18] | Bielderman et al (2013) [19] | King-Kallimanis et al (2014) [17] | Bohn (2022) [29] |
|---|---|---|---|---|
| Data Source | MacArthur Study Waves 1 and 2 (1988-1992) | Local health authority survey of older adults (2008) | SHARE, Wave 1 (2004-2006) | Victoria Longitudinal Study |
| Geography | USA | the Netherlands | 12 EU countries | Canada |
| Participants | Aged 70-79 years | Aged ≥ 65 years | Aged ≥ 50 years | Aged ≥ 53 years |
| *N* | 1010 | 1508 | 27,938 | 649 |
| Frailty Measure | CHS frailty phenotype | GFI frailty phenotype | Frailty phenotype | Multi-morbidity items |
| Number of Items in Measure | 5 | 15 | 7 | 30 |
| Item Types | Continuous and ordinal | Dichotomised | Dichotomised | Ordinal (treated as continuous) |
| Number of Frailty Dimensions | 2 | 3 | 1 | 7 |
| Frailty Dimensions | (1) Physical activity, slowness, weakness; (2) Weight loss and exhaustion | (1) Daily Activities; (2) Psychosocial Functioning; (3) Health Problems | Frailty | (1) Mobility; (2) Instrumental Health; (3) Emotional Wellbeing; (4) Comorbidity; (5) Respiratory Symptoms; (6) Cardiac Symptoms; Physical Activity |
| Frailty pro-files | Not examined | Not examined | Not examined | Not clinically frail; (2) Mobility-type frailty; (3) Respiratory- type frailty |
| Method | PCA | EFA | CFA | EFA |
| Model Fit | Not reported | Not reported | RMSEA <0.045; CFI >0.95 | RMSEA =.03; CFI =.90 |
| Explained Variance | 48% | 50.6% | Not reported | Not reported |
| Association with Out- comes | Regression models for 4-year disability and 9- year mortality | Not examined | Not examined | Risk for accelerated cognitive decline and impairment |
| Drivers Explored | Not examined | Age, education significant. Sex, living situation, financial status not significant. | Not examined | Age not examined. Sex not significant. |

Abbreviations: SHARE, Survey of Health, Ageing and Retirement in Europe; CHS, Cardiovascular Health Study; GFI, Groningen Frailty Indicator; PCA, Principal Component Analysis; EFA, Exploratory Factor Analysis; CFA, Confirmatory Factor Analysis

Table 5: Frailty Indicators Used in the Prior Literature on the Dimensionality of Frailty

| **Fried's Frailty Phenotype [17]** | **Cardiovascular Health Study (CHS) Frailty Criteria [18]** | **Groningen Frailty Indicator [19]** | **British Women's Heart and Health Study (BWHHS) Frailty Indicators [20]** | **Victoria Longitudinal Study (VLS) Frailty Index [21]** | **ELSA Frailty Index** |
|---|---|---|---|---|---|
| Because of health problem expected to last >3 months, do you have difficulty... | Timed 10-foot walk | Shopping | Have you had a fall in past year? | Health has affected ability to travel | Difficulty walking 100m |
| Walking 100 m? | Grip strength | Walking outdoors | Compared with your activity level 3 years ago, are you doing more, same or less? | Health has affected ability to socialize | Difficulty sitting 2 hrs |
| Climbing a flight of stairs without resting? | Energy expenditure–weighted assessments of engagement in recreational, exercise, housework, and yardwork activities | Dressing and undressing | Do you have problems washing or dressing? (no problem, some problem, unable to wash and dress) | Health has affected ability to do hobbies | Difficulty getting up from chair |
| Highest of 4 (2 from each hand) of dynamometer measurement of grip strength in kilograms | Percentage of body weight lost (or gained) between Waves 1 and 2 | Going to the toilet | Is your present state of health causing you problems with household chores? | Health has affected ability to do mental activities | Difficulty climbing several flights of stairs without resting |
| How often do you engage in activities that require a low or moderate level of | During the past week, how much have you been distressed by feeling low in energy | Physical fitness | Difficulty in carrying out activity on their own: going up and downstairs | Health has affected ability to get around town | Difficulty climbing one flight of stairs without resting |
| energy such as gardening, cleaning the car or going for a walk? | or slowed down? | | | | |
| What has your appetite been like? | | Vision problems | Difficulty in carrying out activity on their own: Walking about | Health has affected ability to do chores | Difficulty stooping, kneeling or crouching |
| Have you been eating more or less than usual? | | Hearing problems | Difficulty in carrying out activity on their own: Going out of the house | Bradburn negative affect (restless, lonely, bored, depressed, upset due to criticism) | Difficulty extending arms above shoulders |
| In the past month, have you had too little energy to do the things you want? | | Unintentional weight loss | Do you have trouble with your eyesight? (not simply needing specs) | CES-D “during the past week, my sleep was restless” | Difficulty pulling or pushing large objects |
| | | Use of more than three medicines | Compared to five years ago, is your memory: improved, same, almost as good, worse, much worse? Dementia on medical exam. | CES-D “during the past week, I felt depressed” | Difficulty lifting or carrying weights over 10 pounds |
| | | Memory complaints | Your health over all: are you anxious or depressed, not depressed – moderately, extremely. | CES-D “during the past week, I felt lonely” | Difficulty picking up a 5p coin |
| | | Experience of emptiness | Do you ever have any pain or discomfort in your chest? | Anemia | Difficulty dressing |
| | | Missing people around | Have you ever had a severe pain across the front of your chest lasting for half an hour or more? | Sex-related health problems (i.e., gynecological problems or prostate problems) | Difficulty walking across a room |
| | | Feeling abandoned | Do you usually bring up phlegm (spit) from your chest first thing in the morning in the winter? | Gastrointestinal problems (colitis/diverticulitis, gall bladder trouble, and/or liver trouble) | Difficulty bathing |
| | | Feeling sad/dejected | In the past four years, have you ever had a period of increased cough and phlegm lasting for 3 weeks or more? | Kidney or bladder trouble | Difficulty eating, such as cutting up food |
| | | Feeling nervous/anxious | Do you get short of breath with other people of your own age on level ground? | Feeling short of breath | Difficulty getting in and out of bed |
| | | | Have you ever been told by a doctor that you have or have had asthma? | Bronchitis or emphysema | Difficulty using the toilet |
| | | | Have you ever been told by a doctor that you have or have had bronchitis or emphysema? | Asthma | Difficulty using map |
| | | | Have you ever been told | Pulse pressure | Difficulty preparing a |

| | | | | |
|---|---|---|---|---|
| | | by a doctor that you have or have had arthritis? | | hot meal |
| | | Have you ever been told by a doctor that you have or have had high blood pressure? | Heart trouble | Difficulty shopping for groceries |
| | | Have you ever been told by a doctor that you have or have had thyroid disease? | Hardening of arteries (i.e., atherosclerosis) | Difficulty making phone calls |
| | | Have you ever been told by a doctor that you have or have had a cataract? | High blood pressure | Difficulty taking medications |
| | | Have you ever been told by a doctor that you have or have had glaucoma? | Stroke | Difficulty doing housework / gardening |
| | | Have you ever been told by a doctor that you have or have had depression? | Finger dexterity | Difficulty managing money |
| | | Have you ever been told by a doctor that you have or have had diabetes? | Timed turn | High blood pressure |
| | | Have you ever been told by a doctor that you | Grip strength | Angina |
| | | have or have had gastric or peptic ulcer? | | |
| | | Have you ever been told by a doctor that you have or have had heart attack (MI)? | Use of walker, cane, or wheelchair | Heart attack |
| | | Have you ever been told by a doctor that you have or have had angina? | Stay at home but in chair most of the time | Congestive heart failure |
| | | Have you ever been told by a doctor that you have or have had a stroke? | Health has affected ability to do physical recreational activities | Abnormal heart rhythm |
| | | Have you ever been told by a doctor that you have or have had cancer? | Spinal condition and/or back trouble | Diabetes |
| | | Cardiovascular disease (diagnosed angina, MI, stroke) | Arthritis (rheumatoid and/or osteo) | Stroke |
| | | Body mass index: high or low | | Lung disease |
| | | Postural hypotension: According to consensus definition | | Asthma |
| | | Hypertensive (>140/90) | | Arthritis |
| | | Waist hip ratio (>/<0.85 | | Osteoporosis |
| | | Sinus tachycardia (>100 | | Cancer |

| | | | | | |
|---|---|---|---|---|---|
| | | | bpm) | | Parkinson's<br>Psychiatric condition<br>Alzheimer's<br>Dementia<br>Depressed<br>Felt everything was an effort<br>Restless sleep<br>Lack of happiness<br>Loneliness<br>Lack of life enjoyment<br>Sadness<br>Could not get going much of the time<br>Self-reported general health<br>Eyesight impairment<br>Hearing impairment<br>Fall<br>Hip fracture<br>Joint replacement<br>Pain whilst walking<br>Whether correct day of month given<br>Whether correct month given<br>Whether correct year |
| | | | | | given |
| | | | | | Whether correct day given |
| | | | | | Memory prompt needed |
| | | | | | Whether reaches cutoff for cflisen |
| | | | | | Whether reaches cutoff for cfani |
| | | | | | Whether reaches cutoff for cflisd |
| | | | | | Number of words recalled immediately |
| | | | | | Number of animal mentioned |
| | | | | | Number of words recalled after delay |

Figure 1: Scree Plot

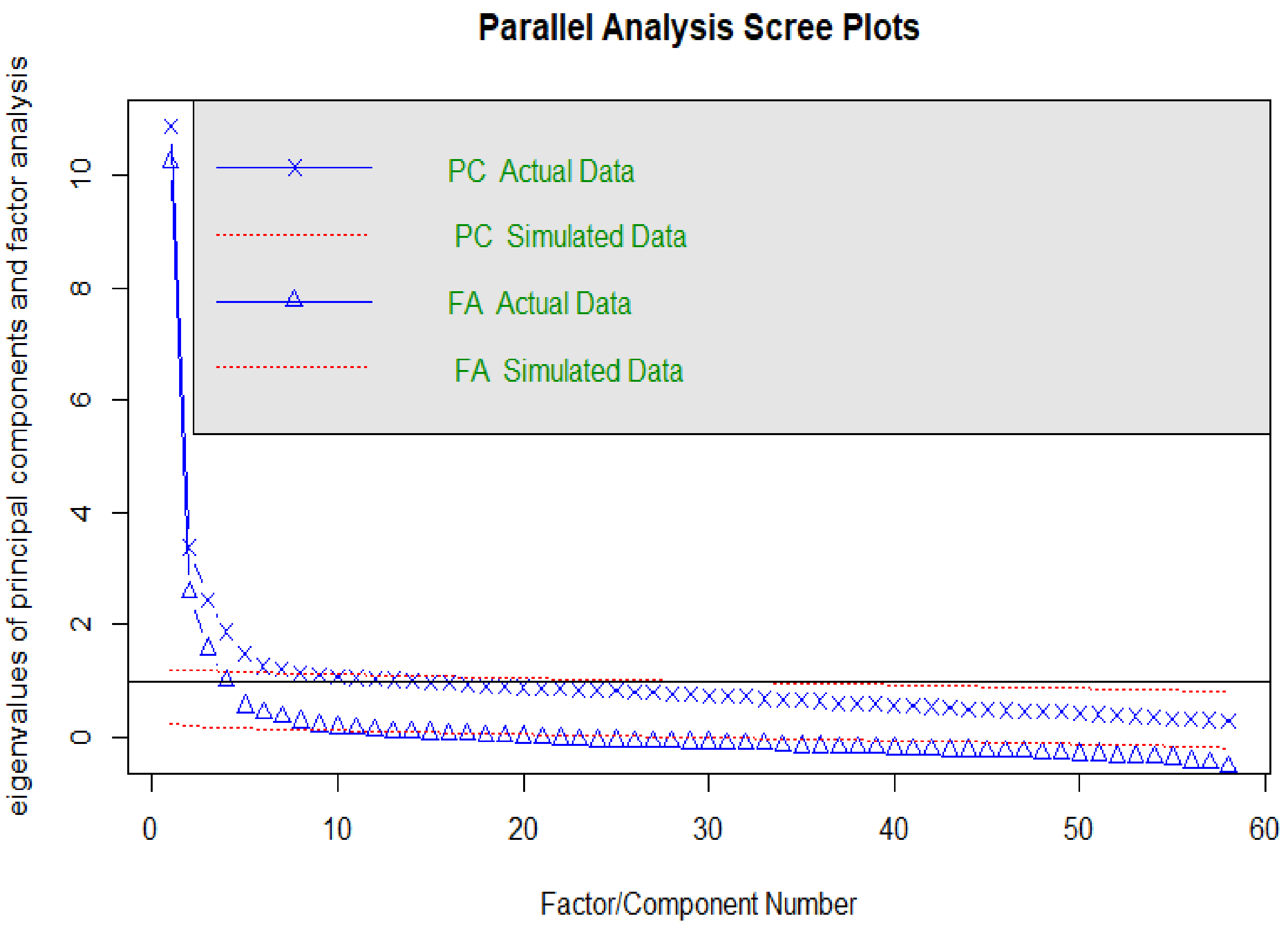

Figure 2. Correlations between Individuals' Latent Factor Scores and Frailty Index Score

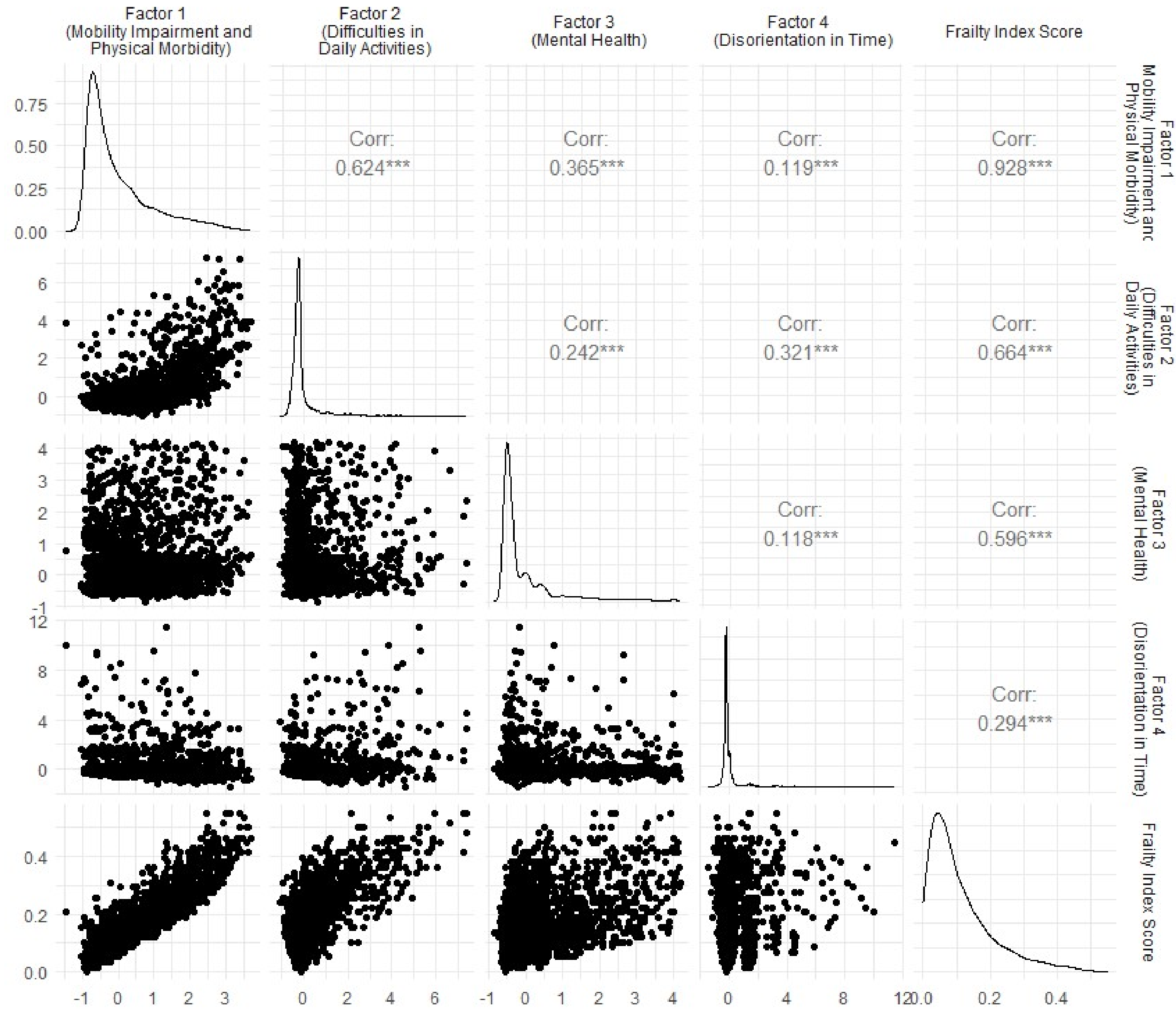